\documentclass[aps,prl,showpacs,twocolumn,floats,epsfig,pdflatex]{revtex4}
\usepackage{epsfig}
\usepackage{amsmath}
\usepackage{amsfonts}
\usepackage{graphicx}
\usepackage{amssymb}
\usepackage{amsbsy}
\usepackage{subfigure}

\begin{document}

\title {Phases and collective modes of Rydberg atoms in an optical lattice}

\author{K. Saha$^{(1)}$, S. Sinha$^{(2)}$, and K. Sengupta$^{(1)}$}
\affiliation{$^{(1)}$ Theoretical Physics Department, Indian
Association for the Cultivation of Science, Jadavpur,
Kolkata-700032, India. \\ $^{(2)}$ Indian Institute of Science
Education and Research-Kolkata - Mohanpur, Nadia 741252, India. }

\date{\today}

\begin{abstract}

We chart out the possible phases of laser driven Rydberg atoms in
the presence of a hypercubic optical lattice. We define a pseudospin
degree of freedom whose up(down) components correspond to the
excited(ground) states of the Rydberg atoms and use them to
demonstrate the realization of a canted Ising antiferromagnetic
(CIAF) Mott phase of the atoms in these systems. We also show that
on lowering the lattice depth, the quantum melting of the CIAF and
density-wave (DW) Mott states (which are also realized in these
systems) leads to supersolid (SS) phases of the atoms. We provide
analytical expressions for the phase boundaries and collective
excitations of these phases in the hardcore limit within mean-field
theory and discuss possible experiments to test our theory.

\end{abstract}

\pacs{03.75.Lm, 05.30.Jp, 05.30.Rt}

\maketitle

The study of ultracold atoms which can be excited to Rydberg states
by suitable laser driving has generated both experimental and
theoretical interest in recent years
\cite{ryd1,ryd2,ryd3,ryd4,ryd5,ryd6,ryd7,ryd8,ryd9,ryd10,ryd11,ryd12}.
Such excited states are known to have large polarizibility which
results in strong Van der Waals ($\sim 1/r^6$) force between them.
This, in turn, leads to the well-known dipole blockade phenomenon
which has been theoretically proposed \cite{ryd1} and experimentally
verified \cite{ryd2}. Further, a collection of such atoms are
predicted to harbor exotic many-body phenomenon such as the presence
of supersolid (SS) droplets within their superfluid (SF) phase
\cite{ryd3} and a second order phase transition between uniform SF
and crystalline SS phases \cite{ryd4,ryd5}. It has also been
conjectured that such systems may act as quantum simulators leading
to realization of qubits \cite{ryd6}. The physics of these atoms in
low-dimensional optical lattices was also shown to lead to several
interesting effects such as the presence of a staircase structure in
the number of excited atoms \cite{ryd7}, presence of ground states
in one-dimension(1D) which hosts non-Abelian excitations such as
Fibonaccci anyons \cite{ryd8}, realization of exotic spin models
with collective fermionic excitations \cite{ryd9}, dynamic creation
of molecular states of such atoms \cite{ryd10}, and realization of
the hard-square model in 2D lattices \cite{ryd11}. The superfluidity
of Rydberg atoms in an 1D optical lattice, confirming the presence
of SF phases in the presence of a lattice in spite of the presence
of the Van der Waals interaction, has also been experimentally
studied \cite{ryd12} . However, the possible phases of Rydberg atoms
in higher dimensional optical lattices has not been charted out so
far.

In this work, we study a system of such Rydberg atoms characterized
by a laser drive frequency $\Omega$, a detuning parameter $\Delta$,
and a Van der Waals interaction strength $V_d$ in the presence of a
hypercubic optical lattice. We present a phase diagram of such a
system and demonstrate the presence of translational symmetry broken
density wave (DW) and canted Ising antiferromagnetic (CIAF)[where
the pseudospin up (down) states correspond to the excited (ground)
states of the Rydberg atoms] Mott phases. We note that such a CIAF
phase amounts to realization of a higher-dimensional translational
symmetry broken spin-ordered ground state using ultracold atoms
\cite{dipole1}. On lowering the lattice depth starting from these
DW/CIAF Mott phases, the atoms undergo successive quantum phase
transitions to SS and SF phases. We provide analytic expressions for
the above-mentioned phase boundaries in the hardcore limit within a
mean-field theory and compute their collective excitations. We point
out that these collective excitations, in the SS phase reached by
increasing hopping strength of the atoms from the CIAF, constitutes
a mixing of the hole-like excitations with the pseudospin collective
modes. This is in contrast to the SS phase obtained analogously from
the DW Mott state, where these modes do not hybridize and thus these
excitations provide a way to distinguish between these two SS
phases. We discuss possible experiments to test our theory. We note
that the properties of ultracold atoms with Rydberg excitations in a
higher dimensional optical lattices has not been studied so far; our
results, particularly the existence of DW, CIAF, and SS phases, are
therefore expected to be of interest to both experimentalists and
theorists working in these fields.

We begin with an effective Hamiltonian of the system, derived within
the rotating wave approximations, given by $H = H_{0} + H_{1} +
H_2$\cite{ryd7}, where
\begin{eqnarray}
H_{0} &=& \Omega \sum_{i} (a_i^{\dagger} b_i +{\rm h.c})
-\mu \sum_i {\hat n}_i + \Delta \sum_i {\hat n}_i^b \nonumber\\
&& + U \sum_i {\hat n}_i^a ({\hat n}_i^a -1) + \lambda U \sum_i
{\hat
n}_i^a {\hat n}_i^b, \nonumber\\
H_1 &=& -J/2 \sum_{\langle ij\rangle} (a_i^{\dagger} a_j + \eta
b_i^{\dagger} b_j + {\rm h.c.}) \nonumber\\
H_2 &=&  V_d/2 \sum_{ij}
({\hat n}_i^b {\hat n}_j^b)/|i-j|^6. \label{ham0}
\end{eqnarray}
Here $a_i(b_i)$ denotes creation operators for the bosons in the
ground(excited) state at the lattice site $i$, ${\hat n}^{a(b)}_i=
a_i^{\dagger} a_i (b_i^{\dagger} b_i)$ are the corresponding number
operators, $\mu$ is the chemical potential, $U(\lambda U)$ is the
on-site interaction strength between two bosons in same (different)
states, $\langle ij\rangle$ indicates that $j$ is one of the
nearest-neighbor sites of $i$, $J(\eta J)$ is the nearest-neighbor
hopping amplitude of the bosons in the ground (excited) states which
can be tuned by tuning the optical lattice depth, and we have set
the lattice spacing to unity. We assume that the Van der Waals
interaction between the Rydberg atoms is strong enough to allow
$n_i^b \le 1$ at each site, but can be neglected for $|i-j| \ge 2$:
$z(z-1)V_d/32 \le \Omega, \Delta, U$, where $z=2d$ denotes the
coordination number of the lattice. In this regime, it is possible
to approximate the long-range interaction term by $H_{2} \simeq
V_d/2 \sum_{\langle ij\rangle}{\hat n}_i^b {\hat n}_j^b$
\cite{comment1}. The simplest variational Gutzwiller wavefunction
which can describe the phases of such a system is given by
$|\psi\rangle = \prod_{i} |\psi\rangle_i$, where
\begin{eqnarray}
|\psi\rangle_i &=& \sum_{n^a_{i},n^b_{i}} f^i_{n^a_{i},n^b_{i}}
|n^a_{i},n^b_{i}\rangle_i, \label{wav}
\end{eqnarray}
and $f^i_{n^a_{i},n^b_{i}}$ are the Gutzwiller coefficients on site
$i$. The variational energy of the system in terms of
$f^i_{n^a_{i},n^b_{i}}$ is given by $E = \langle \psi
|H|\psi\rangle= E_0+E_1+E_2$, where
\begin{eqnarray}
E_0 &=& \sum_{i} \sum_{n^a_{i},n^b_{i}} \Big[ \Big ( -\mu
(n_{i}^a + n_{i}^b) + \Delta n_{i}^b \nonumber\\
&& +\frac{U}{2} [n_{i}^a(n_{i}^a-1) + 2 \lambda n_{i}^a
n_{i}^b]\Big) |f^i_{n^a_{i},n^b_{i}}|^2 \nonumber\\
&& +\Omega \left(\sqrt{n_{i}^a(n_{i}^b+1)}
f_{n^a_{i}-1,n^b_{i}+1}^{i\ast} f^i_{n^a_{i},n^b_{i}} +{\rm h.c.}
\right)\Big ] \nonumber\\
E_1 &=& -J/2 \sum_{\langle ij\rangle} \sum_{n^a_{i,j},n^b_{i,j}}
\Big[ \left( f_{n^a_{i}-1,n^b_{i}}^{i\ast} f_{n^a_{j}
+1,n^b_{j}}^{j\ast}
\sqrt{n_{i}^a(n_{j}^a+1)}\right.\nonumber\\
&& \left.+ \eta f_{n^a_{i},n^b_{i}-1}^{i\ast}
f_{n^a_{j},n^b_{j}+1}^{j\ast}\right) f^i_{n^a_{i},n^b_{i}}
f^j_{n^a_{j},n^b_{j}}  + {\rm h.c} \Big] \nonumber\\
E_2 &=& V_d/2 \sum_{\langle ij\rangle} \sum_{n^a_{i,j},n^b_{i,j}} |
f^i_{n^a_{i},n^b_{i}} f^j_{n^a_{j},n^b_{j}}|^2 n_{i}^b n_{j}^b.
\label{enexp}
\end{eqnarray}
%%%%%%%%%%%%%%%%%%%%%%%%%%%%%%%%%%%%%%%%%%%%%%%%%%%%%%%%%%%%%%%%%%%%
\begin{figure}
\rotatebox{0}{\includegraphics*[width=\linewidth]{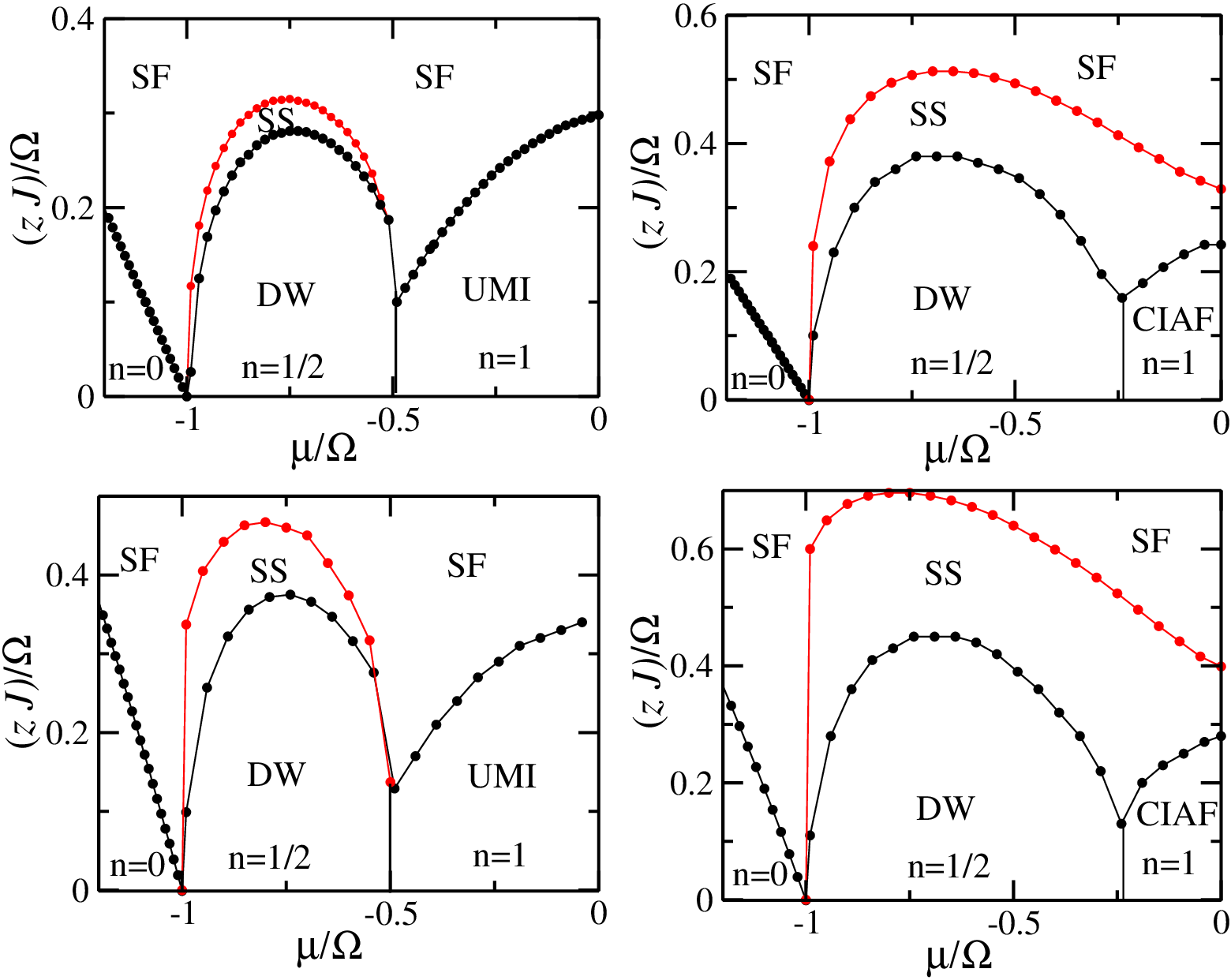}}
\caption{(Color online) Top left(right) panels: $\mu$ vs $J$ phase
diagram for $\mu <0$, $\eta=1$, $\Delta=0$, $U/\Omega=1$,
$\lambda=3$, and $zV_d/\Omega=4(9)$ showing DW, SF, MI, SS and CIAF
phases. Bottom panels: Same as the top panels but with $\eta=0$.}
\label{fig1}
\end{figure}
%%%%%%%%%%%%%%%%%%%%%%%%%%%%%%%%%%%%%%%%%%%%%%%%%%%%%%%%%%%%%%%%%%%%%%%
A numerical minimization $E$ provides the mean-field ground states.
We note that for finite $\Omega$, $H$ conserves $n_i= n^a_i+n^b_i$;
however $n^a_i$ and $n^b_i$ are not conserved. It is evident from
Eq.\ \ref{ham0} that $n_i$ in the ground state is determined by
$\mu$; in contrast, $n_i^b$ for a fixed $\mu$ is determined by a
competition between $\Delta<0$ and $V_d$ which are optimized by
$n_i^b=1$ on every site and $n_i^b=0$ on every alternate site
respectively. This competition provides a possibility of
translational symmetry broken ground states with two sublattice
structure (denoted subsequently as $A$ and $B$). These expectations
are corroborated in the phase diagrams shown in Figs.\ \ref{fig1}
and \ref{fig2} for representative values of the parameter for
$n_i\le 2$ at each site. We find from Fig.\ \ref{fig1} that for $n_i
\le 1$, the Mott phases constitute uniform Mott insulating (MI)
phases with $n_i=0$ and $1$ at each lattice site, and two sublattice
symmetry broken DW and CIAF phases. For the DW phase, the $A$
sublattice has linear combination of $|1,0\rangle$ and $|0,1\rangle$
states where the $B$ sublattice has $n_i=0$ leading to $n_A-n_B=1$
and ${\bar n}=(n_A+n_B)/2=1/2$. The CIAF phase, in contrast has
$n_A=n_B=1$; in this phase the Gutzwiller wavefunction takes the
form
\begin{equation}
|\psi_{A(B)}\rangle = \cos(\theta_{A(B)}) |1,0\rangle -
\sin(\theta_{A(B)}) |0,1\rangle, \label{twomode}
\end{equation}
on $A(B)$ sublattice with the canting angle
$\phi=\theta_A-\theta_B$. The CIAF phase is favored over the uniform
MI phase for $zV_d \ge zV_d^c/\Omega \simeq 8$ as can be seen from
the top left panel of Fig.\ \ref{fig2}. The transition between these
phases is first order. From Fig.\ \ref{fig1}, we also find that upon
increasing $J$, one encounters two second order transitions; the
first occurs from the DW or CIAF phases to a SS phase with $n_A \ne
n_B$ and $\langle b_a \rangle, \langle b_b \rangle \ne 0$ and the
second from the SS to a uniform SF phase. The DW and the SF order
parameters and the CIAF canting angle across the CIAF-SS-SF
transition is shown in right panel of Fig.\ \ref{fig2} for
$zV_d/\Omega=10$ and $\mu/\Omega= 0.2$. We also find the existence
of several multicritical points in the phase diagram where SS, SF,
and DW (left panels of Fig.\ \ref{fig1}) and CIAF, SS, DW, and SF
phases (top right panel of Fig.\ \ref{fig1} and top left panel of
Fig.\ \ref{fig2}) meet. The inclusion of fluctuation may lead to
phase separation near such multicritical points; an analysis of this
effect is beyond the scope of the present mean-field theory. Similar
phase diagram for $\mu>0$, shown in the bottom panel of Fig.\
\ref{fig2}, reveals the existence of $\langle n\rangle =3/2$ DW
phase which has a linear combination of $|1,0\rangle$ and
$|0,1\rangle$ ($|1,1\rangle$ and $|2,0\rangle$) states on $A$($B$)
sublattice. We note from the bottom panels of Fig.\ \ref{fig2} that
the SS phase atop the ${\bar n}=3/2$ DW phase is favored by large
negative $\Delta$.

\begin{figure}
\rotatebox{0}{\includegraphics*[width=\linewidth]{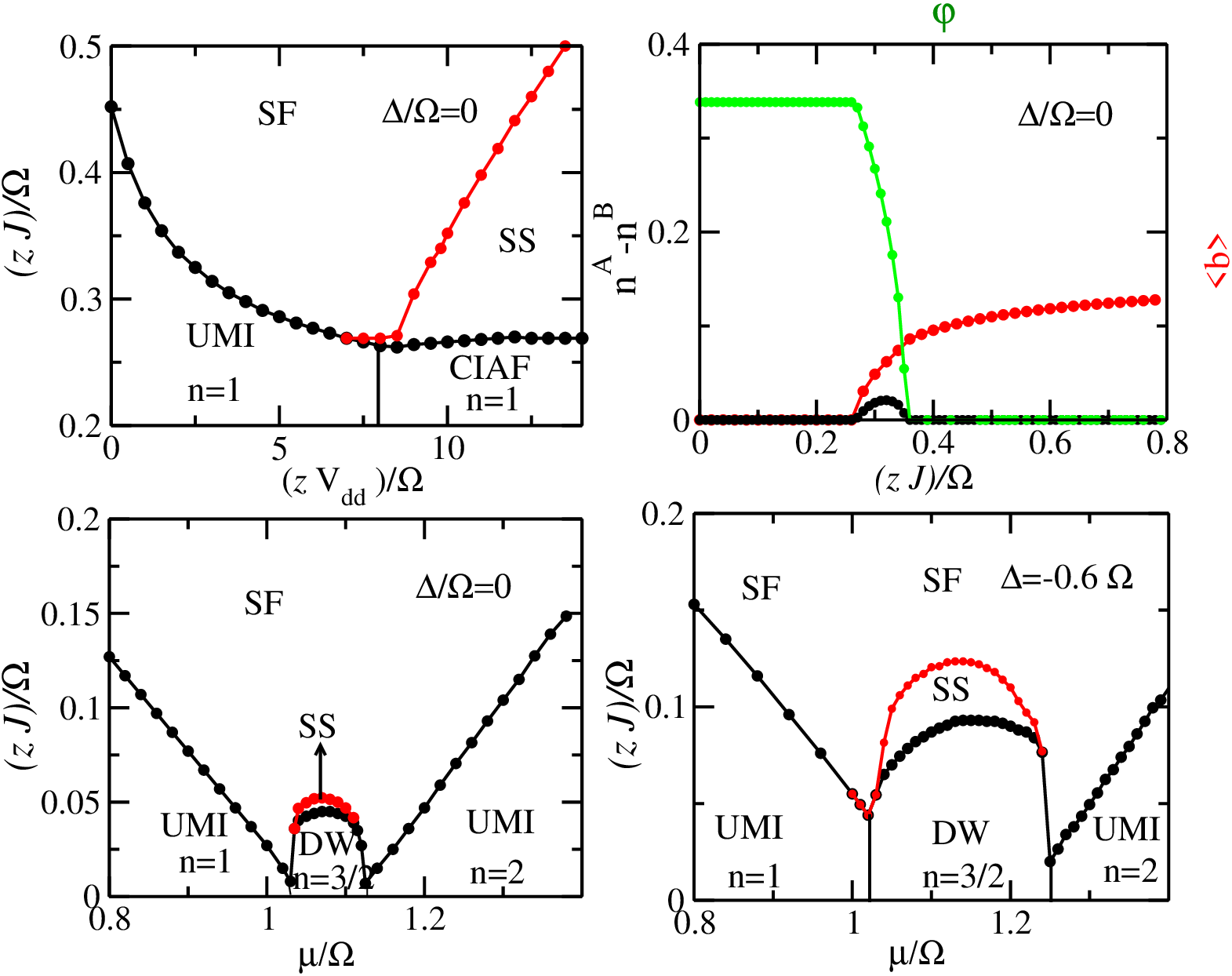}}
\caption{(Color online) Top left panel: Phase diagram as a function
of $V_d$ and $J$ for $\mu/\Omega=0.2$ showing a multicritical point
for $ZV_d/\Omega \simeq 8$ and $ZJ/\Omega \simeq 0.26$. Top right
panel: Plot of the DW (black circle) and SS (red circle) order
parameter and cant angle $\phi$ (green circle) as a function of $J$
for $z V_d/\Omega=10$ and $\mu/\Omega=0.2$ across the CIAF-SS-SF
transition. Bottom panels: $\mu$ vs $J$ phase diagrams for
$zV_d/\Omega=4$ showing the DW with $\langle n\rangle=3/2$, uniform
MI, SS and SF phases. For all plots, $\eta=1$, $\lambda=3.$, and
$U/\Omega=1$.} \label{fig2}
\end{figure}

Next, we consider hardcore bosons ($U \gg \Omega, V_{d},\Delta$)
with filling $n_{i} \leq 1$. In this regime, the single site
wavefunction can be written as $|\psi\rangle_i =
f^{i}_{00}|0,0\rangle + f^{i}_{10}|1,0\rangle +
f^{i}_{01}|0,1\rangle $. To study the dynamics and collective
excitations of the bosons analytically, we generalize Eq.\ \ref{wav}
with time dependent $f^{i}_{n^{a}_{i},n^{b}_{i}}(t)$. The resulting
Schrodinger equations for $f^{i}_{n^{a}_{i},n^{b}_{i}}(t)$ can be
obtained by minimizing the action $S=\int dt \langle \psi|\imath
\partial_{t} - H |\psi \rangle$ with the constraints
$\sum_{n^{a}_{i},n^{b}_{i}}|f^{i}_{n^{a}_{i},n^{b}_{i}}|^2 =1$ for
each site. The ground state phases of the system can also be
obtained as the steady state solutions
$\bar{f}^{i}_{n^a_{i},n^b_{i}}$ of the Schrodinger equation. The
eigenfrequencies of the small fluctuations $\delta \bar{f}^{i}(t)$
around $\bar{f}^{i}_{n^a_{i},n^b_{i}}$ describe the collective
excitations of the quantum phases and determine their
stability\cite{sinha}. The details of these calculations can be
found in Ref.\ \cite{supp1}; here we present the key results
regarding the different phases and their collective modes.

\begin{figure}
\rotatebox{0}{\includegraphics*[width=\linewidth]{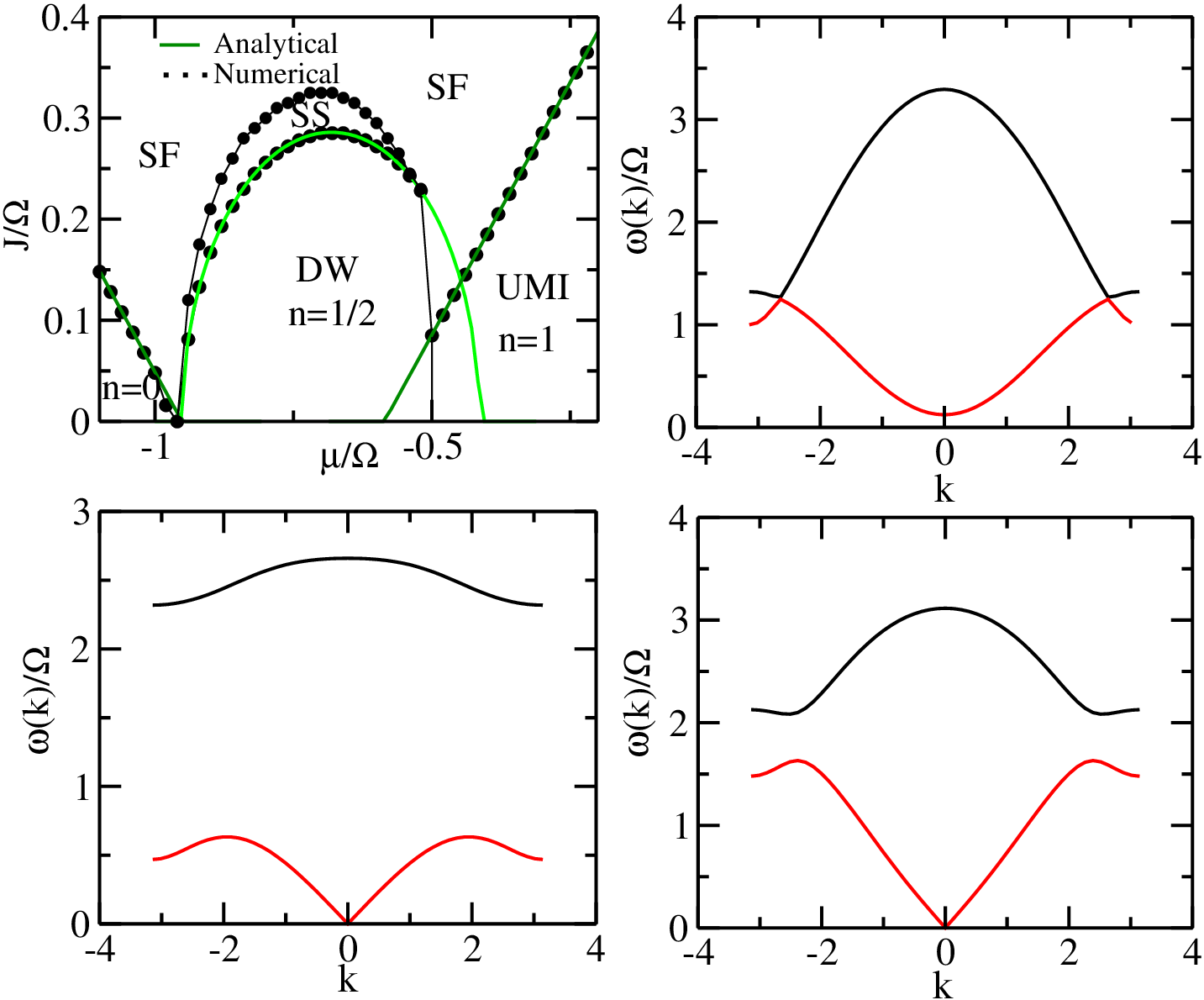}}
\caption{(Color online) Top left panel: Analytical Phase diagram for
$\mu<0$ in the hardcore limit. All parameters except $U$ and
$\lambda$ are same as top left panel of Fig.\ \ref{fig1}. Top right
panel: Excitation spectra of the uniform MI $V/\Omega=1$,
$J/\Omega=\Delta/\Omega=0.1$, $\mu/\Omega=0.2$, $z=6$ and $\eta=1$.
Bottom panels: Excitation spectra of the SF phase over the SS phase
(left panel) and uniform MI phase (right panel). All parameters are
same as those in the top right panel except $J/\Omega=0.1 (0.2)$ and
$\mu/\Omega=-0.75 (0.2)$ for the bottom left (right) panels. Here
$k=|\vec k|$ along $k_x=k_y=k_z$. } \label{fig3}
\end{figure}

{\bf Uniform MI and SF phases}: The uniform MI phase with $\langle
n\rangle=0$ is described by $\prod_i f_{00}^i |0,0\rangle_i$ with
$f^{i}_{00} =1$. For $J=0$, this MI phase appears for $\mu
<-\sqrt{\Omega^2 + \Delta^2/4} + \Delta/2$; with increasing $J$ one
finds a continuous transition to a uniform SF phase. The particle
excitations are described by the fluctuations $\delta f^{i}_{10}$
and $\delta f^{i}_{01}$ with energy,
\begin{eqnarray}
\omega_{\vec k} = \pm \sqrt{\frac{1}{4}\{\Delta + (1
-\eta)\epsilon_{\vec{k}}\}^2 + \Omega^2} + \frac{\Delta}{2} -\mu -(1
+ \eta)\epsilon_{\vec{k}}, \label{particle1} \nonumber\
\end{eqnarray}
where $\epsilon_{\vec{k}} = 2J\sum_{i=1}^{d}\cos(k_{i})$. The phase
boundary between the MI and the SF phases which can be obtained from
the condition $\omega_{\vec k=0} = 0$ is shown in the top left panel
of  Fig.\ \ref{fig3}. For larger values of $\mu$, an uniform MI
phase with $n_i=1$ occurs for $V_{d} < V_d^c$. In this phase the
wavefunction at each site is given by Eq.\ \ref{twomode} with
$\theta_{A}=\theta_{B}$. The collective excitations of this MI phase
is shown in the top right panel Fig.\ \ref{fig3}. For $J>J_c$, the
MI phase becomes unstable by the creation of holes with excitation
energy $\omega^h_{\vec{k}} = -\Omega f_{01}/f_{10} + \mu -J(
|f_{10}|^{2} + \eta |f_{01}|^{2})\epsilon_{\vec k}$ and enters into
a homogeneous SF phase via a second order occurring at
$\omega^h_{\vec k=0}=0$ \cite{transition}. The collective modes of
the SF phase, worked out in details in Ref.\ \cite{supp1}, is shown
the bottom right panel of Fig.\ \ref{fig3} and displays the
well-known massless phase and massive amplitude modes. The pseudo
spin excitation energy of this phase is given by $\omega^{2} =
\Omega^{2}\left[f_{01}^{2}/f_{10}^{2} + f_{10}^{2}/f_{01}^{2} -
2V_{d}f_{10} f_{01} \epsilon_{\vec{k}}/(\Omega J) + 2 \right]$
\cite{supp1}. An instability occurs at a critical strength of Van
der Waals interaction $V_{d}$ for which $\omega =0$ at $\vec{k} =
\pi$. This indicates broken translational symmetry in the uniform
phase and appearance of antiferromagnetic ordering. The phase
diagram so obtained agrees well with the numerical result for the
MI-CIAF transition presented earlier.

\begin{figure}
\rotatebox{0}{\includegraphics*[width=\linewidth]{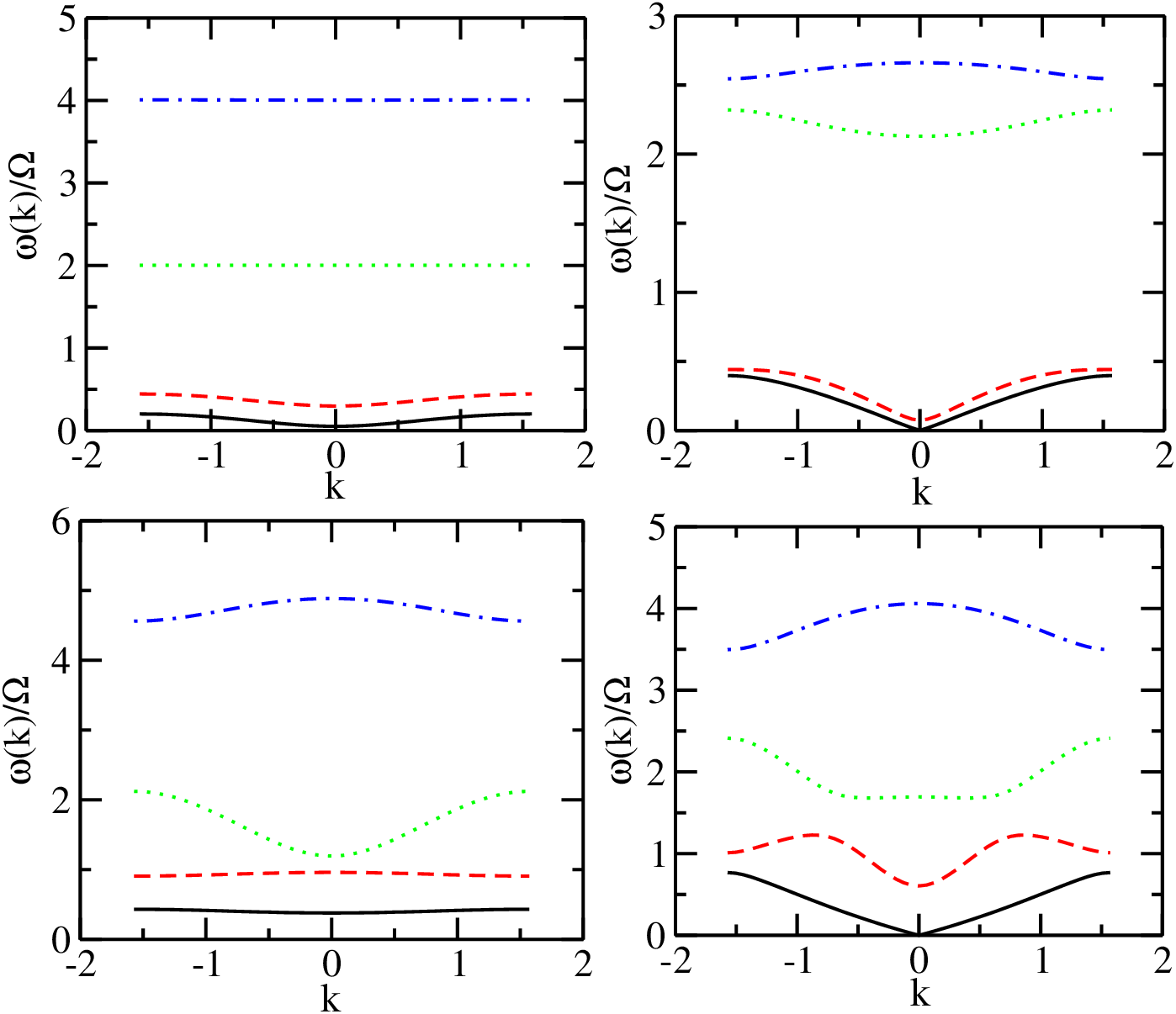}}
\caption{(Color online) Top panels: Excitation spectra of the DW
phase (left) and the SS phase above it (right). The green dotted and
the blue dash-dotted line denote the pseudospin excitation modes
while the black solid and the red dashed lines denote particle/hole
excitation modes. All parameters are same as those in bottom left
panel of Fig.\ \ref{fig3} except $J/\Omega=0.05$ (left) and $0.07$
(right). Bottom panels: Excitation spectra of the CIAF phase (left)
and the SS phase above it (right). All parameters are same as those
in top right panel of Fig.\ \ref{fig3} except $V/\Omega=2$ and
$J/\Omega=0.03$ (left) and $0.15$ (right).} \label{fig4}
\end{figure}

{\bf Density wave with $\langle n\rangle = 1/2$}: The DW state has
only one boson per site on sublattices A whose wavefunction is given
by Eq.\ \ref{twomode} with $\tan 2\theta=2\Omega/\Delta$ and an
empty sublattice $B$. In the hardcore limit, the particles in
sublattice A has an excited state $\sin \theta |1,0\rangle +
\cos\theta |0,1\rangle$ with excitation energy $2\sqrt{\Omega^2 +
\Delta^2/4}$. These can be thought of pseudospin flip excitation.
For $J=0$, another possible excitation is creation of a hole in
sublattice A which costs an energy $E_{h}=\mu + \sqrt{\Omega^2 +
\Delta^2/4} - \Delta/2$. Similarly at sublattice B particle
excitation in two internal states has energy $E_{p\pm}=-\mu +x/2 \pm
\sqrt{x^2/4 + \Omega^2}$ with $x = V_{d}z\sin^2\theta + \Delta$. For
finite $J$, the particle and hole excitations gain dispersions
\cite{supp1} as shown in the top left panel of Fig.\ \ref{fig4};
however, the pseudospin modes remain dispersionless and
well-separated from the particle and hole modes. By increasing $J$,
the DW state enters into a supersolid(SS) phase at $J=J_c$ via a
continuous transition. The phase boundary between the DW and the SS
phases can be obtained analytically by demanding the condition of
one gapless excitation\cite{transition} and is given by
$E_{h}E_{p+}E_{p-}= (zJ)^2[\eta \Omega \sin 2\theta
+(x-\mu)\cos^2\theta - \mu (\eta \sin\theta)^{2}]$. In this SS
phase, the hole-like excitations are well separated in energy from
the pseudospin flip excitations; their dispersion has been derived
in \cite{supp1} and is shown in the top right panel of Fig.\
\ref{fig4}. For larger $\mu$, the DW phase undergoes first order
transitions to SF ($J>J_c$) or MI ($J<J_c$) provided $V_{d} <
V_d^c$. The collective modes of the SF phase is shown in the bottom
left panel of Fig.\ \ref{fig3} and displays the well-known gapless
Goldstone mode. For $V_d >V_d^c$ and $J<J_c$, there is a continuous
transition between the DW and the CIAF phases. The phase diagrams
obtained from this analysis is shown in the top left panel Fig.\
\ref{fig3}.

{\bf CIAF phase:} In this phase each site has $\langle n \rangle=1$
but with different linear combination of the pseudospin up and down
states in the two sublattice leading to a canting angle $\phi$ as
explained earlier. Due to the two sublattice structure, the hole
excitation energies over this ground state are given by \cite{supp1}
\begin{equation}
\omega^{h}_{\pm} = -y_{+}/2 +\mu \pm \left[(f^{A}_{10}f^{B}_{10} +
\eta f^{A}_{01}f^{B}_{01})^{2}\epsilon(k)^{2} +
y_{-}^{2}/4\right]^{1/2} \nonumber\ \label{hole_afm}
\end{equation}
where, $y_{\pm} = \Omega [f^{A}_{01}/f^{A}_{10}\pm
f^{B}_{01}/f^{B}_{10}]$. The CIAF phase melts when
$\omega^h_{-}(k=0) =0$, and a SS phase is formed; the corresponding
phase diagram agrees well with the numerical result plotted in top
left panel of Fig.\ \ref{fig2}. The CIAF phase also has two gapped
pseudospin excitations (bottom left panel of Fig.\ \ref{fig4});
their analytical expressions for general filling $n_{0}$ per site
are given in \cite{supp1}. In contrast to the DW phase, the CIAF
pseudospin modes have finite dispersion and the lower pseudospin
mode is close to the higher hole excitations branch. Consequently,
the SS phase above CIAF, in contrast to its counterpart obtained
from DW, displays an preformed roton like structure resulting from
the hybridization of these hole and pseudospin excitation branches
as depicted in right bottom panel of Fig.\ \ref{fig4}. Thus these SS
phases can be distinguished by their collective mode structure.

%By increasing $J$ further, this modulation in the hole excitation
%leads to the formation of a roton like structure in homogeneous SF
%phase above SS (see bottom right panel of Fig.\ \ref{fig3}).

The experimental verification of the collective modes and the phase
diagram predicted in this work would involve standard experiments
carried out on ultracold atom systems \cite{greiner1,bakr1,ess1}.
Usual momentum distribution measurements would differentiate between
the predicted MI and the SS/SF phases. The DW phase can be
distinguished from the CIAF and the uniform MI phases by the
presence of checkerboard pattern showing odd and even occupation in
alternate sites belonging to different sublattices; such a pattern
can be easily measured in parity of occupation measurement of
individual sites \cite{bakr1}. The distinction between the SS phases
obtained by increasing $J$ starting from CIAF and DW phases would
requires measurement of the dispersion of the collective modes via
lattice modulation or Bragg spectroscopy experiments
\cite{ess1,sk1}. The SS phases would display both checkerboard
pattern for occupation numbers and a momentum distribution peak at
$k=0$ which will distinguish it from other phases.

To conclude, we have charted out the mean-field phase diagram and
computed the collective modes of laser driven Rydberg atoms. Our
work, which is expected to be qualitatively accurate for $d>2$ has
demonstrated the presence of SS, CIAF and DW phases of these atoms
which have distinct collective mode spectra. We note that the CIAF
phase found here constitutes an example of a translation symmetry
broken magnetic ground state in a $d>1$ ultracold atom system.
Possible extension of our work would involve study of effect of
quantum fluctuations on the mean-field phase diagram and a more
detailed incorporation of effect of the dipolar interaction between
the Rydberg atoms which is expected to be important for $V_d \gg
\Omega, \Delta, J, U$. We have suggested several experiments to test
our theory.

\section{Supplementary Material}

To analyze the phases and the collective modes of a system of
Rydberg atoms described by $H$ defined in Eq. 1 of the main text, we
consider the hardcore bosons with $U\rightarrow \infty$. In this
limit the Gutzwiller wavefunction at any site $i$ can be written as
$|\psi\rangle_i = f^{i}_{00}|0,0\rangle + f^{i}_{1,0}|1,0\rangle
+f^{i}_{01}|0,1\rangle$. The mean field energy of the system is
given by:
\begin{eqnarray}
E[\{f^{i}\}] &=&  \Omega \sum_{i} \left[f^{i\ast}_{10}f^{i}_{01} +
f^{i\ast}_{01}f^{i}_{10}\right] -\mu \sum_{i}\left[|f^{i}_{10}|^{2}
\right. \nonumber\\
&& \left. + |f^{i}_{01}|^{2}\right] +\Delta \sum_{i}|f^{i}_{01}|^{2}
+ \frac {V_d}{2}\sum_{\langle ij\rangle}
|f^{i}_{01}|^{2}|f^{j}_{01}|^{2}
\nonumber\\
&& -J\sum_{\langle ij\rangle}
f^{i\ast}_{00}f^{i}_{10}f^{j}_{00}f^{j\ast}_{10} -J\eta
\sum_{\langle
ij\rangle}f^{i\ast}_{00}f^{i}_{01}f^{j}_{00}f^{j\ast}_{01} .
\nonumber\\
\end{eqnarray}
Using time dependent Gutzwiller's wavefunction the action becomes,
\begin{equation}
S= \int dt\left[ i \sum_{i} \{f^{i\ast}_{00}\dot{f}^{i}_{00} +
f^{i\ast}_{10}\dot{f}^{i}_{10}+ f^{i\ast}_{01}\dot{f}^{i}_{01}\}
-E[\{f^{i}\}]\right] \label{action_gutz}
\end{equation}
The Schrodinger equations for $f^{i}(t)$ are given by:
\begin{eqnarray}
i \dot{f}^{i}_{00} & = &
-Jf^{i}_{10}\sum_{\delta}f^{\delta}_{00}f^{\ast \delta}_{10} -J\eta
f^{i}_{01}\sum_{\delta}f^{\delta}_{00}f^{\ast
\delta}_{01}-\lambda_{i}f^{i}_{00}\nonumber\\
i \dot{f}^{i}_{10} & = & \Omega f^{i}_{01} - (\mu +
\lambda_{i})f^{i}_{10} -Jf^{i}_{00}\sum_{\delta}f^{\ast \delta}_{00}
f^{\delta}_{10}.\nonumber\\
i \dot{f}^{i}_{01} & = & \Omega f^{i}_{10} - (\mu - \Delta +
\lambda_{i})f^{i}_{01} + V_{d}f^{i}_{01}
\sum_{\delta}|f^{\delta}_{01}|^{2} \nonumber\\
&& -J\eta f^{i}_{00}\sum_{\delta}f^{\ast
\delta}_{00}f^{\delta}_{01}. \label{dynamic_f}
\end{eqnarray}
where $\delta$ is near neighbor site index of $i^{\rm th}$ site and
$\lambda_{i}$ s are the Lagrange multipliers corresponding to the
wavefunction normalization at each site. To find the frequencies of
the small fluctuations, we decompose each $f^{i}$ in two parts,
$f^{i}(t) = \bar{f}^{i} + \delta f^{i}(t)$. The steady state
solutions of Eq.\ \ref{dynamic_f} corresponding to the ground state
of the system are $\bar{f}^{i}$ and $\delta f^{i}(t)$s are time
dependent small amplitude fluctuations around the steady state
values. We decompose these fluctuations in fourier modes $\delta
f^{j} = e^{-i\omega t} \sum_{k}e^{i \vec{k}. \vec{R}_{j}} \delta
f(\vec{k})$ to obtain the the collective frequencies
$\omega(\vec{k})$ from the linearized dynamical equations [Eq.\
\ref{dynamic_f}]. For the phases with two sublattice structure, we
have used the notation $\bar{f}^{j} = f^{s} + f^{a} e^{i \vec{\pi}.
\vec{R_{j}}}$, and $\lambda_{j} = \lambda_{s} + \lambda_{a}e^{i
\vec{\pi}. \vec{R_{j}}}$, where $\vec{R}_{j}$ is the position of
lattice site $j$.

{\bf {State with $n=0$:}} In this state, $\bar{f}_{00}=1$,
$\bar{f}_{10}=0$, $\bar{f}_{01}=0$. From the steady state solution
of the equation of motion we obtain $\lambda_i=0$ for all sites. The
particle excitations can be obtained from the linearized equations
for $\delta f$
\begin{eqnarray}
\omega\delta f_{10}(\vec{k}) & = & \Omega\delta f_{01}(\vec{k}) -\mu \delta f_{10}(\vec{k}) -\epsilon(\vec{k})\delta f_{10}(\vec{k})\\
\omega\delta f_{01}(\vec{k}) & = & \Omega\delta f_{10}(\vec{k})
-(\mu -\Delta)\delta f_{01}(\vec{k}) -\eta\epsilon(\vec{k})\delta
f_{01}(\vec{k}) \nonumber\\
\end{eqnarray}
which leads to the particle excitations with two internal degrees
given by,
\begin{eqnarray}
\omega_{k} &=& \pm \sqrt{\{\Delta + (1 -\eta)\epsilon(\vec{k})\}^2/4
+ \Omega^2} \nonumber\\
&& + \frac{\Delta}{2} -\mu -(1 + \eta)\epsilon(\vec{k})/2,
\end{eqnarray}
where, $\epsilon(\vec{k}) = 2J\sum_{i=1}^{d}\cos(k_{i})$. The
instability of this phase takes place for $k =0$ leading to the
formation of homogeneous SF phase. The phase boundary is given by:
\begin{equation}
(\mu  + Jz)(\mu - \Delta + J\eta z) = \Omega^{2}
\end{equation}

{\bf DW state with $n=1/2$:} This DW state has two sublattice
structure and the wavefunction is given by $|1,0,1,0...\rangle$. The
sites of B sublattice are empty and $f^{B}_{00}=1$. The particles at
sublattice A are in linear superposition of ground state and Rydberg
state, with $f^{A}_{10} = \cos \theta$ and $f^{A}_{01} = -\sin
\theta$. The minimization of $E[\{f^i\}]$ gives $\tan 2\theta =
2\Omega/\Delta$. From the steady state condition obtained from
equating the right side of Eq.\ \ref{dynamic_f} to zero, we can fix
the Lagrange multipliers to be $\lambda^{A} = -\mu +\Delta/2
-\sqrt{\Delta^{2}/4 + \Omega^{2}}$ and $\lambda^{B} =0$. In momentum
space the linearized equations for fluctuations can thus be written
as,
\begin{eqnarray}
-\omega \delta f^{\ast +}_{00}(\vec{k}) &=&   -2f^{S}_{10}\epsilon(\vec{k})\delta f^{-}_{10}(\vec{k}) -2\eta f^{S}_{01}\epsilon(\vec{k})\delta f^{-}_{01}(\vec{k}) \nonumber\\
&& -2\lambda_{S}\delta f^{\ast +}_{00}(\vec{k})\nonumber\\
\omega\delta f^{-}_{10}(\vec{k}) &=&  \Omega\delta
f^{-}_{01}(\vec{k}) -\mu \delta f^{-}_{10}(\vec{k}) -2f^{S}_{10}
\epsilon(k)\delta f^{\ast +}_{00}(\vec{k})\nonumber\\
\omega\delta f^{-}_{01}(\vec{k}) &=&  \Omega\delta
f^{-}_{10}(\vec{k}) -(\mu -\Delta )\delta f^{-}_{01}(\vec{k})
\nonumber\\
&& + 4V_{d}z|f^{S}_{01}|^{2}\delta f^{-}_{01}(\vec{k})  - 2\eta
f^{S}_{01} \epsilon(k) \delta f^{\ast +}_{00}(\vec{k}), \label{eigeq}\nonumber\\
\end{eqnarray}
where $\delta f^{\pm}(\vec{k}) = \delta f(\vec{k}) \pm \delta
f(\vec{k} + \vec{\pi})$ and $\lambda_S= (\lambda^A+\lambda^B)/2$. We
note that $\delta f^{i \ast}$ satisfy similar equations with
$\omega$ replaced by $-\omega$. In the atomic limit, for $J=0$ we
obtain the particle(hole) excitations of sublattice B(A)
analytically from the above equations. Removing a particle from
sublattice A (hole excitation) costs an energy $E_{h}=\mu +
\sqrt{\Omega^2 + \Delta^2/4} - \Delta/2$. This is the eigenvalue of
fluctuations $\delta f^{+}_{00}$. The particle excitations in
sublattice B can be obtained by diagonalizing the single site atomic
hamiltonian written in the basis of $|10\rangle$ and $|01\rangle$
states. Particle excitation in two internal states has energy
$E_{p\pm}=-\mu +x/2 \pm \sqrt{x^2/4 + \Omega^2}$ with $x =
4V_{d}zf^{S2}_{01} + \Delta$. These are the eigenvalues
corresponding to the fluctuations $f^{-}_{01}$ and $f^{-}_{10}$ in
Eq.\ \ref{eigeq}. For non vanishing $J$ the eigenvalues can be
obtained by numerically solving the cubic equation. For a second
order transition to SS phase, the phase boundary can be obtained
from the condition of vanishing eigenvalue at $k=0$ and the
analytical expression can be obtained from the determinant of the
eigenvalue equations which can be read off from Eq.\ \ref{eigeq},
\begin{eqnarray}
2\lambda_{S}\left[\mu (\mu -x) -\Omega^{2}\right]&=& 4
J^{2}z^{2}\left[2 \eta \Omega f^{S}_{10}f^{S}_{01} +
\eta^{2}f^{S2}_{01} \mu  \right.  \nonumber\\
&& \left. + f^{S2}_{10}(\mu -x)\right] \label{phbound_dw}
\end{eqnarray}
where $x = 4V_{d}zf^{S2}_{01} + \Delta$, . From the expressions of
$E_{p\pm}$ and $E_h$ written down earlier and using $f_{10} =
\cos(\theta)$ and $f_{01}=-\sin(\theta)$, it can be shown that this
condition is equivalent to $E_{h}E_{p+}E_{p-}= (zJ)^2[\eta \Omega
\sin 2\theta +(x-\mu)\cos^2\theta - \mu (\eta \sin\theta)^{2}]$
which is used in the main text.

{\bf Mott phase with $n=1$}: In this phase each site contains
exactly one particle, which is a linear superposition of the ground
state and excited state. If we represent ground state by
$|\downarrow\rangle$ and excited state by $|\uparrow\rangle$, then
an effective quantum spin Hamiltonian can written for this state:
\begin{eqnarray}
H_{spin} &=& \Omega\sum_{i}
\left[|\uparrow\rangle\langle\downarrow|_{i} +
|\downarrow\rangle\langle\uparrow|_{i}\right] + \sum_{\langle
ij\rangle} P_{i} V_{ij} P_{j} \nonumber\\
&& +\Delta\sum_{i} |\uparrow\rangle\langle\uparrow|_{i}.
\end{eqnarray}
where $P = |\uparrow\rangle\langle\uparrow|$.

{\bf Uniform phase:} In this phase, one has at each site, $f_{00} =
0$, $f_{10} = \cos\theta$, $f_{01} = \sin\theta$, and $\theta$ can
be obtained by minimizing the energy,
\begin{equation}
E/N = \Omega \sin 2\theta + \frac{V_{d}z}{2} \sin^{4} \theta +
\Delta \sin^{2}\theta
\end{equation}
The Lagrange multiplier is given by $\lambda = \Omega f_{01}/f_{10}
- \mu$. The excitation energy corresponding to the fluctuation
$\delta f_{00}$  is given by,
\begin{equation}
\omega \delta f_{00}(\vec{k}) = -\left[ (|f_{10}|^{2} +\eta
|f_{01}|^{2})\epsilon(\vec{k}) + \lambda \right] \delta
f_{00}(\vec{k})
\end{equation}
The uniform Mott insulator to SF transition takes place due to the
instability at $k = 0$ at $Jz(|f_{10}|^{2} + \eta |f_{01}|^{2})= \mu
-\Omega f_{01}/f_{10}$. The spin excitations can be obtained from
the linearized equations,
\begin{eqnarray}
\omega \delta f_{10}(\vec{k}) &= & \Omega\delta f_{01}(\vec{k}) -(\mu + \lambda)\delta f_{10}(\vec{k}) \\
\omega\delta f_{01}(\vec{k})  &=&  \Omega\delta f_{10}(\vec{k}) -(\mu -\Delta + \lambda)\delta f_{01}(\vec{k}) \nonumber\\
&& +V_{d}z |f_{01}|^{2}\delta f_{01}(\vec{k}) +V_{d}
f_{01}^{2}(\epsilon(\vec{k})/J) \nonumber\\
&& \times (\delta f_{01}(\vec{k}) + \delta f_{01}^{\ast}(\vec{k}))
\end{eqnarray}
The energy of these excitations can be easily calculated to yield
\begin{equation}
\omega^{2} = \Omega^{2}\left[\frac{f_{01}^{2}}{f_{10}^{2}} +
\frac{f_{10}^{2}}{f_{01}^{2}} - \frac{2V_{d}f_{10} f_{01}
\epsilon(\vec{k})}{\Omega  J} + 2 \right] \label{mott_sw}
\end{equation}
{\bf Canted Ising antiferromagnetic phase with $n=1$} In this phase
each site has one particle but this phase has antiferromagnetic
order. In this phase we have two sublattice values of $f_{10}$ and
$f_{01}$. Fluctuation of $f_{00}$ is given by,
\begin{eqnarray}
\omega\delta f_{00}(\vec{k}) & = & -\beta \epsilon(\vec{k})\delta
f_{00}(\vec{k})
-\lambda_{s}\delta f_{00}(\vec{k}) \nonumber\\
&& -\lambda_{a}\delta f_{00}(\vec{k}+\vec{\pi})\\
\omega\delta f_{00}(\vec{k}+\vec{\pi})& = & \beta
\epsilon(\vec{k})\delta f_{00}(\vec{k}+\vec{\pi}) -\lambda_{s}\delta
f_{00}(\vec{k}+\vec{\pi}) \nonumber\\
&& -\lambda_{a}\delta f_{00}(\vec{k}).
\end{eqnarray}
with, $ \beta = (|f^{s}_{10}|^{2} - |f^{a}_{10}|^{2}) + \eta
(|f^{s}_{01}|^{2} - |f^{a}_{01}|^{2}) $. The Lagrange multipliers
are $\lambda_{s}=\Omega (f^{A}_{01}/f^{A}_{10} +
f^{B}_{01}/f^{B}_{10})/2 - \mu$ and $ \lambda_{a}=\Omega
(f^{A}_{01}/f^{A}_{10} - f^{B}_{01}/f^{B}_{10})/2$. The hole
excitation energy is given by,
\begin{equation}
\omega^{h}_{\pm} = -\lambda_{s} \pm
\sqrt{\beta^{2}\epsilon(\vec{k})^{2} + \lambda_{a}^{2}}.
\label{cifahole}
\end{equation}
A straightforward substitution for $\beta$, $\lambda_s$ and
$\lambda_a$ in Eq.\ \ref{cifahole} leads to Eq. (6) used in the main
text.

The phase boundary can be obtained from the instability of above
excitation at $k = 0$, $ \beta^{2}z^{2} = \lambda_{s}^{2} -
\lambda_{a}^{2}$. The spin modes can be obtained from the
fluctuations $\delta f_{10}$ and $\delta f_{01}$,
\begin{eqnarray}
\omega \delta f_{10}(\vec{k}) &=& \Omega\delta f_{01}(\vec{k}) -(\mu + \lambda_{s})\delta f_{10}(\vec{k}) \nonumber\\
&& -\lambda_{a} \delta f_{10}(\vec{k}+\vec{\pi})\\
\omega\delta f_{01}(\vec{k}) &=&  \Omega\delta f_{10}(\vec{k}) -(\mu
-\Delta + \lambda_{s}) \delta f_{01}(\vec{k})
\nonumber\\
&& -\lambda_{s}\delta f_{01}(\vec{k} + \vec{\pi})
+V_{d}z (|f^{s}_{01}|^{2} + |f^{a}_{01}|^{2})\nonumber\\
&& \times \delta f_{01}(\vec{k})
-V_{d}z(2f^{s}_{01}f^{a}_{01})\delta f_{01}(\vec{k}+\vec{\pi})
\nonumber\\
&& +V_{d} (|f^{s}_{01}|^{2} - |f^{a}_{01}|^{2})(\epsilon(\vec{k})/J)
\nonumber\\
&& \times (\delta f_{01}(\vec{k}) + \delta f_{01}^{\star}(\vec{k})).
\end{eqnarray}
A similar set of equations can be obtained for $\delta f(\vec{k} +
\vec{\pi})$ (replacing $\vec{k} \rightarrow \vec{k} + \vec{\pi}$ in
above equations) and for $\delta f^{\ast}$ (taking complex conjugate
of above equations and replacing $\omega$ by $-\omega$).

The spin modes can also be obtained for a CIAF with general filling
$\langle n\rangle = n_{0}$ from a simple variational calculation.
The variational wave function at $i$th site can be written as,
\begin{equation}
|\psi\rangle_{i} = \cos \theta_{i}|n_{0},0\rangle + e^{i\phi_{i}}
\sin \theta_{i}|n_{0}-1,1\rangle.
\end{equation}
The Lagrangian is given by,
\begin{eqnarray}
{\it L} &=&  \sum_{i}\left[\dot{\phi}_{i}\sin^{2}\theta_{i} + \Omega \sqrt{n_{0}}\sin 2\theta_{i} \cos \phi_{i} \right. \nonumber\\
&& \left. + U(n_{0} -1)(\cos^{2}\theta_{i} + \lambda \sin^{2} \theta_{i}) \right]\nonumber\\
&& + \frac{V_{d}}{2} \sum_{i\ne j} \sin^{2}\theta_{i}
\sin^{2}\theta_{j} + \frac{U}{2}(n_{0}-1)(n_{0} -2) .
\end{eqnarray}
From variation of the Lagrangian we obtain following equations,
\begin{eqnarray}
& &\dot{\theta}_{i} = -\Omega \sqrt{n_{0}} \sin\phi_{i}\\
& & \sin2\theta_{i}\dot{\phi}_{i} + 2\Omega
\sqrt{n_{0}}\cos2\theta_{i} \cos{\phi}_{i} + V_{d} \sin2\theta_{i}
\sum_{\delta} \sin^{2}\theta_{i} \nonumber\\ & & + U (\lambda
-1)(n_{0} -1)\sin2\theta_{i}=0.
\end{eqnarray}
After linearization and eliminating $\delta \phi$, we obtain
\begin{eqnarray}
\omega^{2} \delta \theta_{k} &=& \beta_{1} \delta \theta_{k} +
\gamma_{+} \epsilon(\vec{k}) \delta \theta_{k} + \beta_{2} \delta
\theta_{k + \pi} \nonumber\\
&& + \gamma_{-} \epsilon(\vec{k}) \delta \theta_{k + \pi}  \\
\omega^{2} \delta \theta_{k + \pi} &=& \beta_{1} \delta \theta_{k +
\pi} - \gamma_{+} \epsilon(\vec{k}) \delta \theta_{k + \pi} +
\beta_{2}
\delta \theta_{k } \nonumber\\
&& - \gamma_{-} \epsilon(\vec{k}) \delta \theta_{k},
\end{eqnarray}
where,
\begin{eqnarray}
\beta_{1} & = & 2 \Omega^{2}n_{0}\left[2 + \cot^{2}2\theta_{A} + \cot^{2}2\theta_{B} \right] \nonumber\\
\beta_{2} & = & 2 \Omega^{2}n_{0}\left[\cot^{2}2\theta_{A} - \cot^{2}2\theta_{B} \right] \nonumber\\
\gamma_{\pm} & = & -\Omega \sqrt{n_{0}} V_{d} (\sin2\theta_{A}  \pm
\sin2\theta_{B})/(2J).
\end{eqnarray}
The excitation energy of spin modes are thus given by,
\begin{equation}
\omega^{2} = \beta_{1}  \pm \sqrt{ \beta_{2}^{2} + (\gamma_{+}^{2} -
\gamma_{-}^{2}) \epsilon_{k}^{2}}.
\end{equation}

This completes our derivation of the excitation spectra of the
different phases of the hardcore bosons. We would like to note that
we expect the qualitative nature of the spectra to remain unchanged
for large $U/\Omega\gg 1$ where the particle modes with $n_i >1$,
neglected in the hardcore limit, are gapped out and do not play an
essential role in determining the low-energy properties of the
system.

\begin{thebibliography}{99}
\bibitem{comment1} We expect the phases obtained to remain same
qualitatively even when this condition is midly violated; see Ref.\
\cite{ryd9} for details.
\bibitem{transition} The analytical phase boundary obtained from
this condition is valid only for continuous transition. This results
in a mismatch of the phase boundaries between numerics and analytics
at the phase boundary of the DW and uniform MI phases around
$\mu/\Omega=-0.5$ as seen in top left panel of Fig.\ \ref{fig3}.
\end{thebibliography}
\end{document}